\documentclass[reprint,aps,prb,amsmath,amssymb]{revtex4-1}

\usepackage[]{graphicx}
\usepackage[]{units}
\usepackage{times}
\usepackage{bm}
\usepackage[ulem=normalem]{changes}
\usepackage[normalem]{ulem}		
\usepackage{color}
\usepackage{xr} 				
\usepackage{cancel}
\definechangesauthor[color=blue]{SM}

\newcommand{\im}{\Im\textnormal{m}}
\newcommand{\comment}[1]{}

\renewcommand{\emph}{\textit}

\newcommand{\one}{1\!\!1}

\begin{document}

\title{Theory of exciton dynamics in time-resolved ARPES: intra- and intervalley scattering in two-dimensional semiconductors}
\author{Dominik Christiansen$^{1}$}
\author{Malte Selig$^{1}$}
\author{Ermin Malic$^{2}$}
\author{Ralph Ernstorfer$^{3}$}
\author{Andreas Knorr$^{1}$}
\affiliation{$^{1}$Institut f\"ur Theoretische Physik, Nichtlineare Optik und Quantenelektronik, Technische Universit\"at Berlin,  10623 Berlin, Germany}
\affiliation{$^{2}$Chalmers University of Technology, Department of Physics, SE-412 96 Gothenburg, Sweden}
\affiliation{$^{3}$Fritz-Haber-Institut der Max-Planck-Gesellschaft, Faradayweg 4-6, 14195 Berlin, Germany}

\begin{abstract}
Time- and angle-resolved photoemission spectroscopy (trARPES) is a powerful spectroscopic method to measure the ultrafast electron dynamics directly in momentum-space. However, band gap materials with exceptional strong Coulomb interaction such as monolayer transition metal dichlacogenides (TMDC) exhibit tightly bound excitons, which dominate their optical properties. This rises the question whether excitons, in particular their formation and relaxation dynamics, can be detected in photoemission. Here, we develope a fully microscopic theory of the temporal dynamics of excitonic time- and angle resolved photoemission with particular focus on the phonon-mediated thermalization of optically excited excitons to momentum-forbidden dark exciton states. We find that trARPES is able to probe the ultrafast exciton formation and relaxation throughout the Brillouin zone.
\end{abstract}

\maketitle


\section{Introduction}

We develop a theoretical description of time- and angle-resolved two-photon photoemission (trARPES) signals to evaluate its potential to address the temporal dynamics of Coulomb- and phonon-induced effects on the optically excited electron and exciton dynamics. As an exemplary material, we use two-dimensional transition metal dichalcogenides (TMDC), which exhibit remarkable electronic and optical properties including a direct band gap at the $K$- and $K'$- points lying at the edges of the hexagonal Brillouin zone \cite{Mak2010,Splendiani2010}. As atomically thin semiconductors, TMDCs possess a reduced dielectric screening of the Coulomb interaction, compared to the bulk case, that gives rise to the formation of a variety of excitons with binding energies of hundreds of meV \cite{Ramasub2012,He2014,Chernikov2014,Wang2018,Mueller2018}. Because of a complex electronic quasi-particle band structure \cite{Steinhoff2015,Kormanyos2015,Rasmussen2015,Sanchez2016}, TMDCs possess a variety of optically addressable bright excitonic states as well as momentum- \cite{MacDonald2015,Louie2015,Selig2016,Ermin2018} and spin-forbidden \cite{MacDonald2015,Louie2015,Echeverry2016} dark excitonic states. In order to study the relaxation dynamics in this complicated excitonic landscape, different experimental techniques such as optical pump-probe \cite{Schmidt2016,Steinleitner2018}, luminescence spectroscopy \cite{Zhang2015,Lindlau2017,Lindlau2018} and time- and angle-resolved photoemission spectroscopy \cite{Bertoni2016,Wallauer2016,Puppin2019} have been performed. The advantage of trARPES over pure optical experiments involving solely transitions between valence and conduction band is that in the latter many possible excitation and relaxation pathways contribute to the measured signal, which makes the identification of the major electronic processes difficult. Time-resolved ARPES, however, possesses a momentum resolution enabling an imaging of the Coulomb correlated electron dynamics of an optically excited state directly in the momentum-space \cite{Damascelli2003,Toben2005,Gudde2007,Suzuki2009,Wang2015}. In this context, for materials like monolayer TMDCs, with optical properties dominated by excitons, the question arises whether trARPES is able to discriminate between excitons as bound electron-hole pair states or electron-hole scattering (free) pair states and whether it can follow the exciton dynamics. In particular, first recent theoretical studies \cite{Perfetto2016,Steinhoff2017,Rustagi2018} suggest that trARPES signals arise from the ionization of excitons: The corresponding signal is located below the conduction band minimum reflecting the excitonic binding energy. However, so far, no description of exciton scattering dynamics, including phonon-induced formation and thermalization observed in trARPES is available.

In this article, based on a many particle Hamiltonian (Sec. \ref{sec:Hamilton}) and the Heisenberg equation of motion formalism (Sec. \ref{sec:EoM}), we present a fully time- and angle resolved microscopic study describing the impact of Coulomb interaction between electrons and holes to the trARPES signal in two-dimensional semiconductor structures, such as TMDCs, after optical excitation. In extension to previous studies \cite{Rustagi2018,Rustagi2019}, we explicitly include not only bright, optically excitable excitons, but also recently introduced momentum-forbidden dark excitonic states \cite{Selig2018} that are generated by the temporally resolved thermalization dynamics due to exciton-phonon scattering and contribute to the optical line shape of TMDCs \cite{Selig2016,Christiansen2017}. This includes $K\Lambda$ and $KK'$-excitons with a hole at the $K$-point and an electron at the $\Lambda$-point or $K'$-point, respectively. After optical excitation, we find first a trARPES imaging of the excitonic coherence and observe the suceeding formation of incoherent, scattering induced, excitonic signals (Sec. \ref{sec:ARPES}). Our theoretical calculations reveal a method to determine the time scales of exciton formation and relaxation with a direct access to momentum-forbidden dark excitonic states \cite{Selig2018}.

\section{Hamiltonian} \label{sec:Hamilton}
\begin{figure}
\begin{center}
\includegraphics[scale=0.8]{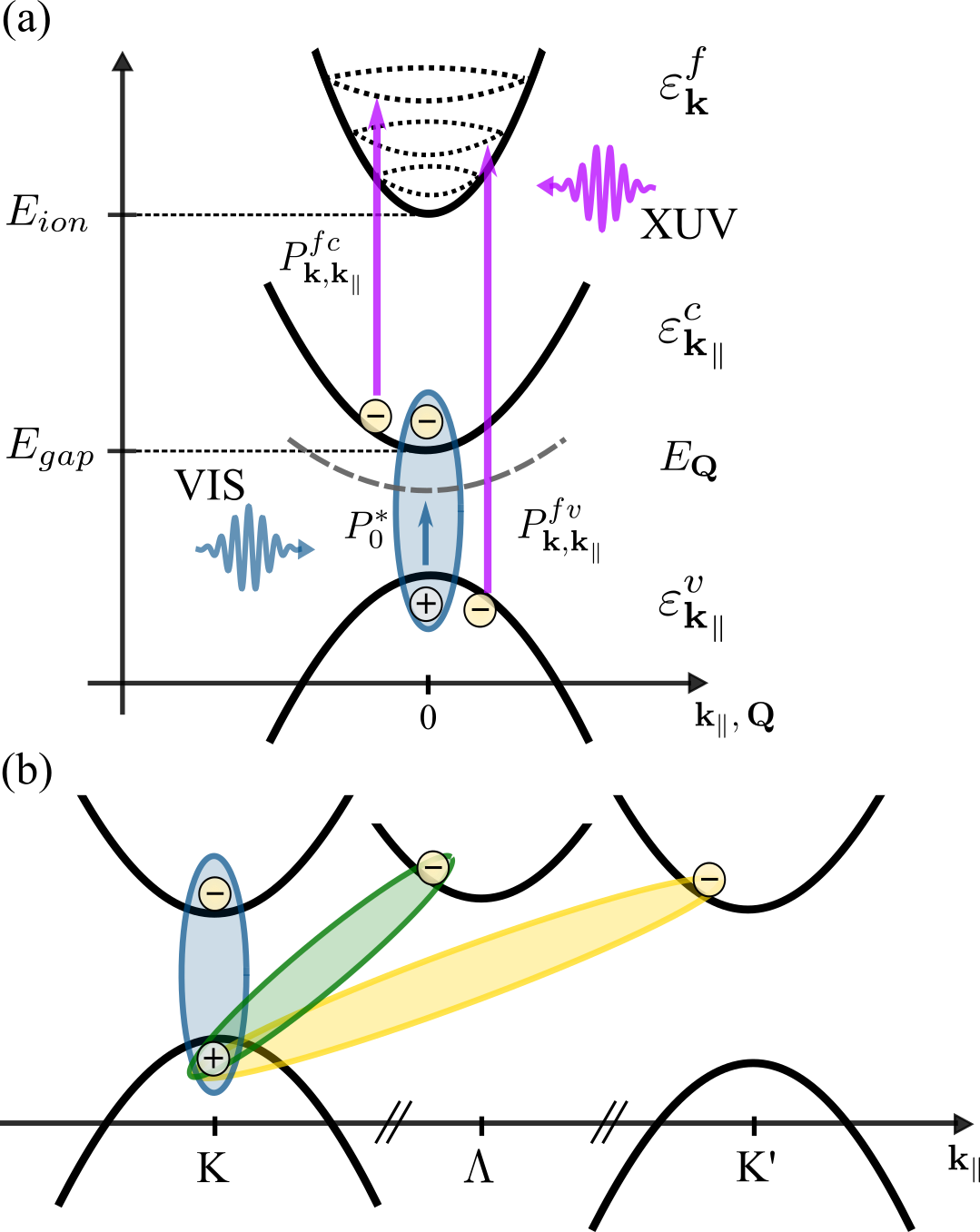}
\end{center}
\caption{(a) First the pump pulse excites a coherent exciton, second the probe pulse simultaneously raises a valence and conduction band electron in the free continuum states, described by a manifold of parabolas. During the transition into the vacuum the in-plane momentum is conserved. (b) Different possible exciton states considering the complex electronic band structure, including the bright $KK$-exciton and momentum-indirect $K\Lambda$ or $KK'$-exciton.}
\label{fig:sketch}
\end{figure}

The theoretical description of the process of the two-photon photoemission (2PPE) \citep{Timm2004,Sakaue2005,Ueba2007,Zeiser2005,Wolf1999,Weinelt2002} consists of two interfering, partly simultaneously occuring subprocesses, cf. Fig. \ref{fig:sketch} (a): First, the optical excitation with a visible (VIS) pump pulse close to the band edge generates correlated electron-hole pairs in the atomically thin TMDC layer (two-dimensional electronic band structure for valence and conduction band), followed by the escape of electrons to the vacuum (three-dimensional dispersion) due to the incidence of an extreme ultraviolet (XUV) probe pulse. Because of the electron-hole interaction, the first optical pump pulse generates tightly-bound electron-hole pairs -- excitons (exciton momentum $\mathbf{Q}$ dispersion $E_{\mu,\mathbf{Q}}$ of state $\mu$ as dashed line). Therefore, the optical preparation and the subsequent dynamics contains valuable information about the excitonic properties and the time dynamics in the semiconductor. Figure \ref{fig:sketch} (a) shows the two fold excitation scheme via the transition amplitudes $P^{\xi_v\xi_c}_{\mu,\mathbf{Q}=0}$ for the VIS (blue) and $P^{cf\xi_c}_{\mathbf{k_{\parallel},k}}$ and $P^{vf\xi_v}_{\mathbf{k_{\parallel},k}}$ for the XUV (purple) excitation. Here, only excitons with wave number $\mathbf{Q}=\mathbf{k}_{\parallel}^c-\mathbf{k}_{\parallel}^v=0$ can be optically excited by the VIS excitation. After that, excitons with $\mathbf{Q}\neq 0$, in particular incoherent excitons, can only be generated by further electron-phonon scattering events (see below). Important to note is that the photon energy of the VIS pulse is smaller compared to the work function of the material, in contrast to the XUV pulse. Therefore, the seperation of different electronic transitions excited by pump and probe pulse is ensured by their different photon energies.

To describe the trARPES signal we discuss the two contributions to the Hamiltonian: (i) the many band TMDC part, involving the optical VIS pump pulse and (ii) the emission process of TMDC electrons into vacuum states, involving the XUV probe pulse.

(i) The many-particle Hamiltonian for the TMDC contributions (index 1) is given by:
\begin{align}
H^{(1)}=H_0^{(1)}+H_{field}^{(1)}+H_{c-c}^{(1)}+H_{ph}^{(1)}+H_{el-ph}^{(1)}.
\end{align}
The TMDC band structure contribution $H_0^{(1)}$:
\begin{align}
H_0^{(1)}&=\sum_{\xi_c,\mathbf{k}_{\parallel}}\varepsilon^{c\xi_c}_{\mathbf{k}_{\parallel}}~c^{\dagger\xi_c}_{\mathbf{k}_{\parallel}}c^{\xi_c}_{\mathbf{k}_{\parallel}}+\sum_{\xi_v,\mathbf{k}_{\parallel}}\varepsilon^{v\xi_v}_{\mathbf{k}_{\parallel}}~v^{\dagger\xi_v}_{\mathbf{k}_{\parallel}}v^{\xi_v}_{\mathbf{k}_{\parallel}}
\end{align}
contains the single-particle energies $\varepsilon_{\mathbf{k}_{\parallel}}^{v/c \xi_{v/c}}$ for valence $(v)$ and conduction $(c)$ band electrons treated in effective mass approximation in the vicinity of each high symmetry point \cite{Dresselhaus1956,Elliott1957,Kormanyos2015}. The fully occupied valence band is assumed to lie at \unit[0]{eV} and the conduction band minimum is energetically situated at the free-particle band gap. Here, $c^{(\dagger)\xi_c}_{\mathbf{k}_{\parallel}}$ and $v^{(\dagger)\xi_v}_{\mathbf{k}_{\parallel}}$ annihilate (create) a conduction- and valence electron in the valley $\xi_c$, $\xi_v$ with wave number $\mathbf{k}_{\parallel}$, respectively. We explicitly include the high symmetry points $K$, $K'$, $\Lambda$ and $\Lambda'$, cp. Fig \ref{fig:sketch} (b). Note that the wave vector $\mathbf{k}_{\parallel}$ is defined with respect to the corresponding valley and the total wave vector is obtained by adding the valley wave vector $\mathbf{k}_{\parallel}+\boldsymbol{\xi}_{c/v}$. The spin is assumed to be equal for all electrons. In our analysis, we neglect the lower valence band at the $K$-valley as well as the $\Gamma$-valley due to the large energetic separation from the upper valence band in monolayer TMDCs \cite{Kormanyos2015}. The electronic excitations of the atomically thin TMDC material are confined in the $x$-$y$-plane ($\mathbf{k}_{\parallel}$). The light-matter interaction reads:
\begin{align}
H_{field}^{(1)}&=\sum_{\mathbf{k}_{\parallel},\xi_c,\xi_v }\mathbf{d}^{cv\xi_c\xi_v}_{\mathbf{k}_{\parallel}} \cdot \mathbf{E}_{vis}(t)~c^{\dagger\xi_c}_{\mathbf{k}_{\parallel}}v^{\xi_v}_{\mathbf{k}_{\parallel}} \delta_{\xi_c,\xi_v}+\text{h.c}.
\end{align}
The optical VIS pulse $\mathbf{E}_{vis}(t)$ generating electron-hole excitations is treated semi-classically in $\mathbf{r}\cdot\mathbf{E}$ - coupling and acts as a source of the single-particle interband transition $P^{\xi_v\xi_c}_{\mathbf{k}_{\parallel}}=\langle v^{\dagger\xi_v}_{\mathbf{k}_{\parallel}}c^{\xi_c}_{\mathbf{k}_{\parallel}}\rangle$ between valence and conduction band. Thereby, $\mathbf{d}^{cv\xi_c\xi_v}_{\mathbf{k}}$ describes the dipole transition matrix element and $\mathbf{E}_{vis}(t)$ denotes the optical pump pulse. For the optically induced interband transition only wave number vertical transitions occur ($\mathbf{k}^v_{\parallel}=\mathbf{k}^c_{\parallel}$ and $\xi_v=\xi_c$) and the sum over the valley indices is restricted to the $K$- and $K'$-point, considering an optical near gap excitation of the TMDC monolayer. The Coulomb interaction between electrons and holes is included by:
\begin{align}
H_{c-c}^{(1)}&=\frac{1}{2}\sum_{\substack{\xi_c,\xi_v, \\ \mathbf{k_{\parallel},k_{\parallel}',q}}}V^{vc\xi_v\xi_c}_{\mathbf{k}_{\parallel},\mathbf{k}_{\parallel}',\mathbf{q}}~c^{\dagger\xi_c}_{\mathbf{k_{\parallel}+q}} v^{\dagger\xi_v}_{\mathbf{k_{\parallel}'-q}} v^{\xi_v}_{\mathbf{k_{\parallel}'}} c^{\xi_c}_{\mathbf{k}_{\parallel}}+\text{h.c}
\end{align}
with the matrix element $V^{vc\xi_v\xi_c}_{\mathbf{k_{\parallel},k_{\parallel}',q}}=V_{\mathbf{q}}\int d^3r \int d^3r'~\psi^{*v\xi_v}_{\mathbf{k}_{\parallel}'}(\mathbf{r})\psi^{*c\xi_c}_{\mathbf{k}_{\parallel}}(\mathbf{r}')e^{i\mathbf{q(r-r')}}\psi^{v\xi_v}_{\mathbf{k}_{\parallel}'}(\mathbf{r})\psi^{c\xi_c}_{\mathbf{k}_{\parallel}}(\mathbf{r}')$ \cite{Kochbuch,Hakenbuch}. Here, the Coulomb potential is treated by an analytical model of the dielectric function within a dielectric environment, including the non-linear $\mathbf{q}$-dependency, beyond the Rytova-Keldysh approximation\cite{Trolle2017,Rytova1967,Keldysh1978,Berkelbach2013}. Since TMDC excitons are spectrally stable over a wide doping and excitation range \cite{Steinhoff2018,Scharf2019}, a static treatment of the Coulomb potential captures well all excitonic effects in the low-density regime. This is different to metals, where excitonic excitations are built up on top of an interacting electron gas \cite{Cui2014,Silkin2015}. The details on the introduced electronic wave functions $\psi^{v/c\xi_{v/c}}_{\mathbf{k}_{\parallel}}$, the transition dipole element and the Coulomb matrix element are discussed in the appendix. To include dissipation, i.e. incoherent exciton dynamics such as exciton density formation and relaxation \cite{Selig2018}, we include the electron-phonon interaction in the TMDC layer:
\begin{align}
H_{ph}^{(1)}+H_{el-ph}^{(1)}&=\sum_{\xi,\alpha,\mathbf{K}}\hbar\Omega^{\xi\alpha}_{\mathbf{K}}~b^{\dagger\alpha\xi}_{\mathbf{K}}b^{\alpha\xi}_{\mathbf{K}} \nonumber \\
&\hspace{-22mm}+\sum_{\substack{\xi_c,\xi_c',\\ \alpha,\mathbf{k_{\parallel},K}}} \left(g^{c\xi_c\xi_c'\alpha}_{\mathbf{K}} c^{\dagger\xi_c}_{\mathbf{k_{\parallel}+K}}c^{\xi_c'}_{\mathbf{k}_{\parallel}}\right)\left(b^{\alpha\xi_{c}-\xi_{c}'}_{\mathbf{K}}+b^{\dagger\alpha\xi_{c}'-\xi_{c}}_{-\mathbf{K}}\right) \nonumber \\
&\hspace{-22mm}+\sum_{\substack{\xi_v,\xi_v',\\ \alpha,\mathbf{k_{\parallel},K}}}\left(g^{v\xi_v\xi_v'\alpha}_{\mathbf{K}} v^{\dagger\xi_v}_{\mathbf{k_{\parallel}+K}}v^{\xi_v'}_{\mathbf{k}_{\parallel}}\right) \left(b^{\alpha\xi_{v}-\xi_{v}'}_{\mathbf{K}}+b^{\dagger\alpha\xi_{v}'-\xi_{v}}_{-\mathbf{K}}\right).
\end{align}
Here, $b^{(\dagger)\alpha\xi}_{\mathbf{K}}$ denotes the annihilation (creation) of a phonon with mode $\alpha$ at the valley $\xi$ and two-dimensional wave vector $\mathbf{K}$. $g^{c/v\xi_{c/v}\xi_{c/v}'\alpha}_{\mathbf{K}}$ is the electron-phonon matrix element for electronic transitions from the valley $\xi_{c/v}'$ to the valley $\xi_{c/v}$ in the conduction or valence band. The Hamiltonian describes electrons scattering from $\xi_{c/v}'+\mathbf{k}_{\parallel}$ to $\xi_{c/v}+\mathbf{k_{\parallel}+K}$ under absorption (emission) of phonons with wave vector $\xi_{c/v}-\xi_{c/v}'+\mathbf{K}$ ($-\xi_{c/v}+\xi_{c/v}'-\mathbf{K}$). The electron-phonon interaction takes into account intra- and intervalley scattering of electrons with an effective deformation potential approximation. We include two acoustic (LA, TA) and two optical phonon modes (LO, TO), which show the strongest deformation coupling to electrons \cite{Jin2014,Li2013,Kaasbjerg2012,Kaasbjerg2013}.

(ii) The next contribution to the Hamiltonian describes the emission process of Coulomb-correlated electrons into the vacuum initiated by a XUV pulse. In the following, $\varepsilon^f_{\mathbf{k}}$ denotes the dispersion of the free electron continuum above the ionization threshold $E_{ion}$ of the TMDC \cite{Guo2016,Keyshar2017}. The transitions from the semiconductor to the vacuum states are induced by a XUV probe pulse $\mathbf{E}_{xuv}(t)$. We introduce the electron operators $f^{(\dagger)}_{\mathbf{k}}$ annihilating (creating) an electron in the three-dimensional continuum states of the vacuum:
\begin{align}
H_0^{(2)}&=\sum_{\mathbf{k}}\varepsilon^{f}_{\mathbf{k}}~f^{\dagger}_{\mathbf{k}}f^{\mathstrut}_{\mathbf{k}} \\
H_{field}^{(2)}&=\sum_{\xi_c,\mathbf{k'}}\mathbf{d}^{cf\xi_c}_{\mathbf{k}'} ~ \delta_{\mathbf{k}_{\parallel},\mathbf{k}_{\parallel}'}\cdot\mathbf{E}_{xuv}(t) ~ c^{\dagger\xi_c}_{\mathbf{k}_{\parallel}}f^{\mathstrut}_{\mathbf{k}'} \nonumber \\
&+\sum_{\xi_v,\mathbf{k'}}\mathbf{d}^{vf\xi_v}_{\mathbf{k}'} ~ \delta_{\mathbf{k}_{\parallel},\mathbf{k}_{\parallel}'}\cdot\mathbf{E}_{xuv}(t) ~ v^{\dagger\xi_v}_{\mathbf{k}_{\parallel}}f^{\mathstrut}_{\mathbf{k}'}+\text{h.c} .
\end{align}
The continuum of the vacuum states is described by a three-dimensional wave vector $\mathbf{k}\in\mathbb{R}^3$. In the course of electron spectroscopy, the probe pulse excites electrons from the valence and conduction band to the vacuum states via the transition amplitudes $P^{vf\xi_v}_{\mathbf{k,k}_{\parallel}}=\langle v^{\dagger\xi_v}_{\mathbf{k}_{\parallel}} f^{\mathstrut}_{\mathbf{k}}\rangle$ and $P^{cf\xi_c}_{\mathbf{k,k}_{\parallel}}=\langle c^{\dagger\xi_c}_{\mathbf{k}_{\parallel}} f^{\mathstrut}_{\mathbf{k}}\rangle$, respectively. Here, the dipole matrix reads $\mathbf{d}^{c/vf\xi_{c/v}}_{\mathbf{k}}=-ie\int d^3r~\psi^{*c/v\xi_{c/v}}_{\mathbf{k}_{\parallel}}\nabla^{\mathstrut}_{\mathbf{k}}\psi^f_{\mathbf{k}}$, where $\mathbf{k=(k_{\parallel}},k_z)$. The conservation of the in-plane momentum $\delta_{\mathbf{k_{\parallel},k_{\parallel}'}}$ follows directly from the optical matrix element. From the emitted photoelectron distribution $\rho^f_{\mathbf{k}}=\langle f^{\dagger}_{\mathbf{k}}f^{\mathstrut}_{\mathbf{k}}\rangle$, due to the conservation of the in-plane wave vector, we obtain information about the wave number distribution of valence and conduction band electrons inside the TMDC material.

\section{Dynamical equations} \label{sec:EoM}

In trARPES experiments the detector measures the photocurrent of electrons, which are emitted by the probe pulse in a certain direction sensitive to their kinetic energy. Using the Heisenberg equation of motion, we develop a description of the time- and angle-resolved photoemitted vacuum electron distribution \citep{Freericks2009,Ramakrishna2004,Eckstein2008,Braun2015} as a function of energy $\varepsilon^f_{\mathbf{k}}$ and in-plane wave number $\mathbf{k}_{\parallel}$:
\begin{align}
I_{\mathbf{k}_{\parallel},\varepsilon^f_{(\mathbf{k}_{\parallel},k_z)}}(\tau)=\lim_{t\rightarrow\infty} \int_{-\infty}^t dt' ~ \partial_{t'}\rho^f_{\mathbf{k}}(t',\tau) , \label{eq:observable}
\end{align}
where $\rho_{\mathbf{k}}^f(t,\tau)$ is the vacuum electron density depending on real time $t$ and the time delay $\tau$ between the VIS and XUV pulse. Since the time integral extends to $\infty$ we count all photoelectrons, which reach the detector. In principle, to derive the two photon photoemission signal, a detector model must be considered. Here, defining Eq. \eqref{eq:observable} as the signal we assume a detector
sensitive to the final state occupation as the 2PPE signal \cite{Weinelt2002}. We note that the time-integrated final state occupation has been used as alternative definition of the photoemission signal assuming a detector sensitive to the total electron number \cite{Petek1997,Gumhalter2018}. We expect however, at least for our calculations, equally informative results for both limiting situations. The occuring situation is similar to problems in the definition of non-stationary signals via photon detection \cite{Kira1999} with simultaneous energy and time resolution.

In the following, for all equations the rapid carrier frequency pulse oscillation contribution has been split off in a rotating wave approximation for each pulse. For the vacuum electron distribution, which determines the observable in Eq. \eqref{eq:observable}, we find:
\begin{align}
\frac{d}{dt}\rho^f_{\mathbf{k}}&=-2\im\left(\Omega^{vf\xi_v}_{\mathbf{k}} P^{vf\xi_v}_{\mathbf{k_{\parallel},k}}+\Omega^{cf\xi_c}_{\mathbf{k}}P^{cf\xi_c}_{\mathbf{k_{\parallel},k}}\right) , \label{eq:rhof}
\end{align}
where we defined the Rabi frequency $\Omega^{c/vf\xi_{c/v}}_{\mathbf{k}}(t)=d^{c/vf\xi_{c/v}}_{\mathbf{k}}\cdot E_{xuv}(t)/\hbar$ and $E_{xuv}(t)$ as XUV pulse envelope. Equation \ref{eq:rhof} shows that the source of the vacuum electrons are the transition amplitudes $P^{cf\xi_c}_{\mathbf{k,k}_{\parallel}}$ and $P^{vf\xi_v}_{\mathbf{k,k}_{\parallel}}$ between conduction/valence band to the vacuum, referred to as photoemission process, cp Fig. \ref{fig:sketch} (a).

In order to take into account the electron-hole Coulomb coupling for the conduction band electrons in Eq. \eqref{eq:rhof} we insert a unit operator $\one=|0\rangle\langle 0|+\sum_{\xi_v,\mathbf{k}^v_{\parallel}}v^{\xi_v}_{\mathbf{k}^v_{\parallel}}v^{\dagger\xi_v}_{\mathbf{k}^v_{\parallel}}+\sum_{\xi_c,\mathbf{k}^c_{\parallel}}c^{\dagger\xi_c}_{\mathbf{k}^c_{\parallel}}c^{\xi_c}_{\mathbf{k}^c_{\parallel}}+\mathcal{O}(na_B^2)$ exploiting the completness relation of the Fock space \cite{Usui1960,Ivanov1993,Schafer1994,Katsch2018}. $n$ denotes the pair (surface) density and $a_B$ the exciton Bohr radius. Higher-order contributions to the unit operator are neglected since we restrict ourself to the low-density regime. $|0\rangle$ denotes the ground states of the semiconductor with completly filled valence and empty conduction band. The expansion of $c^{\dagger\xi_c}_{\mathbf{k}_{\parallel}}f^{\mathstrut}_{\mathbf{k}}$ in Eq. \eqref{eq:rhof}, by inserting the unit opreator, yields:
\begin{align}
c^{\dagger\xi_c}_{\mathbf{k}_{\parallel}}f^{\mathstrut}_{\mathbf{k}}=\sum_{\xi_v,\mathbf{k}^v_{\parallel}} c^{\dagger\xi_c}_{\mathbf{k}_{\parallel}}v^{\xi_v}_{\mathbf{k}^v_{\parallel}}v^{\dagger\xi_v}_{\mathbf{k}^v_{\parallel}}f^{\mathstrut}_{\mathbf{k}} .
\end{align}
Since we consider the optical excitation of undoped semiconductors below the free-particle band gap only the second contribution to the unit operator is relevant. With this procedure the conduction electron operators are expressed uniquely by electron-hole pair operators. To treat the quantum mechanical hierarchy problem arising from the many-particle interaction we exploit the cluster expansion scheme \cite{Axt1994,Lindberg1994,Fricke1996}. Analogous the valence band electrons in $v^{\dagger\xi_v}_{\mathbf{k}_{\parallel}}f^{\mathstrut}_{\mathbf{k}}$ of Eq. \eqref{eq:rhof} need to be expanded, which yields $v^{\dagger\xi_v}_{\mathbf{k}_{\parallel}}f^{\mathstrut}_{\mathbf{k}}=\sum_{\xi_c,\mathbf{k}^c_{\parallel}} v^{\dagger\xi_v}_{\mathbf{k}_{\parallel}}c^{\dagger\xi_c}_{\mathbf{k}^c_{\parallel}}c^{\xi_c}_{\mathbf{k}^c_{\parallel}}f^{\mathstrut}_{\mathbf{k}}=v^{\dagger\xi_v}_{\mathbf{k}_{\parallel}}f^{\mathstrut}_{\mathbf{k}}-\sum_{\xi_c,\mathbf{k}^c_{\parallel}} v^{\dagger\xi_v}_{\mathbf{k}_{\parallel}}c^{\xi_c}_{\mathbf{k}^c_{\parallel}}c^{\dagger\xi_c}_{\mathbf{k}^c_{\parallel}}f^{\mathstrut}_{\mathbf{k}}$. Here, the interband Coulomb interaction leads to corrections to the dominating $v^{\dagger\xi_v}_{\mathbf{k}_{\parallel}}f^{\mathstrut}_{\mathbf{k}}$ term. Since we restrict our analysis to the leading order justified by a weak optical excitation of the sample such that the valence band occupation $\rho^{v\xi_v}_{\mathbf{k}_{\parallel}}\approx 1$ holds for all investigated scenarios, we can neglect the Coulomb-induced contribution here. For the vacuum electron density $\rho^f_{\mathbf{k}}$ including Coulomb coupling we find:
\begin{widetext}
\begin{align}
\frac{d}{dt} \rho^f_{\mathbf{k}}&=-2\im\left(\Omega^{vf\xi_v}_{\mathbf{k}} P^{vf\xi_v}_{\mathbf{k_{\parallel},k}} + \Omega^{cf\xi_c}_{\mathbf{k}} P^{*\xi_c\xi_v}_{\mathbf{k}_{\parallel}}P^{vf\xi_v}_{\mathbf{k}_{\parallel},\mathbf{k}} ~ e^{-\frac{1}{i\hbar}\varepsilon_{vis}t} + \sum_{\xi_v,\mathbf{k}^v_{\parallel}} \Omega^{cf\xi_c}_{\mathbf{k}} \delta\langle P^{\dagger\xi_c\xi_v}_{\mathbf{k},\mathbf{k}^v_{\parallel}}P^{vf\xi_v}_{\mathbf{k}^v_{\parallel},\mathbf{k}}\rangle \right) . \label{eq:rhof_cc}
\end{align}
\end{widetext}
The first term accounts for the photoemission of valence band electrons. The second term stems from interband Coulomb interaction in Hartree-Fock limit and couples the interband transition with the transition between valence band and vacuum. The third term is a correlated two-particle quantity $\delta\langle c^{\dagger\xi_c}_{\mathbf{k_{\parallel}+q}}v^{\xi_v}_{\mathbf{k}^v_{\parallel}+\mathbf{q}}v^{\dagger\xi_v}_{\mathbf{k}^v_{\parallel}}f^{\mathstrut}_{\mathbf{k}}\rangle$, describing the Coulomb correlated photoemission, obtained beyond the Hartree-Fock limit $\delta\langle a^{\dagger}_1a^{\dagger}_2a^{\mathstrut}_3a^{\mathstrut}_4\rangle=\langle a^{\dagger}_1a^{\mathstrut}_3\rangle\langle a^{\dagger}_2a^{\mathstrut}_4\rangle-\langle a^{\dagger}_1a^{\mathstrut}_4\rangle\langle a^{\dagger}_2a^{\mathstrut}_3\rangle+\langle a^{\dagger}_1a^{\dagger}_2a^{\mathstrut}_3a^{\mathstrut}_4\rangle$. While in the lowest Hartree-Fock level all correlations between the carriers are neglected, in the first order the appearing correlated quantity contains the true two-body interaction describing deviations from the factorization. This term runs over the valence band electrons, cf. Eq. \eqref{eq:rhof_cc}. The kinetics of the photoemission of valence band electrons reads:
\begin{align}
\frac{d}{dt}P^{vf\xi_v}_{\mathbf{k_{\parallel},k}}&=\frac{1}{i\hbar}\left(\varepsilon^f_{\mathbf{k}}-\varepsilon^{v\xi_v}_{\mathbf{k}_{\parallel}}-\varepsilon_{xuv}\right)P^{vf\xi_v}_{\mathbf{k_{\parallel},k}} +\partial_t P^{vf\xi_c}_{\mathbf{k_{\parallel},k}}\vert_{scatt} \nonumber \\
&-i\Omega^{fv\xi_v}_{\mathbf{k}} \rho^{v\xi_v}_{\mathbf{k}_{\parallel}} -i\Omega^{fc \xi_c}_{\mathbf{k}}P^{\xi_v\xi_c}_{\mathbf{k}_{\parallel}} ~ e^{\frac{1}{i\hbar}\varepsilon_{vis} t}. \label{eq:Pvf_eh}
\end{align}
The solution Eq. \eqref{eq:Pvf_eh} oscillates with the kinetic energy of the vacuum and valence band electrons and is driven by the electronic valence band occupation. Schematically, using the notation $\partial_t P^{vf\xi_c}_{\mathbf{k_{\parallel},k}}\vert_{scatt}$, we include phonons, which lead to a dephasing and a broadening $\gamma$ of the transition.

The equation of motion of the correlated two-particle quantity reads:
\begin{widetext}
\begin{align}
\frac{d}{dt} \delta\langle P^{\dagger\xi_c\xi_v}_{\mathbf{k}_{\parallel},\mathbf{k}^v_{\parallel}} P^{vf\xi_v}_{\mathbf{k}_{\parallel}^v,\mathbf{k}} \rangle &= \frac{1}{i\hbar}\left(\varepsilon^f_{\mathbf{k}}+\varepsilon^{v\xi_v}_{\mathbf{k}^v_{\parallel}}-\varepsilon^{v\xi_v}_{\mathbf{k}^v_{\parallel}}-\varepsilon^{c\xi_c}_{\mathbf{k}_{\parallel}}-\varepsilon_{xuv}\right) \delta\langle P^{\dagger\xi_c\xi_v}_{\mathbf{k}_{\parallel},\mathbf{k}_{\parallel}^v} P^{vf\xi_v}_{\mathbf{k}^v_{\parallel},\mathbf{k}} \rangle +\partial_t \delta\langle P^{\dagger\xi_c\xi_v}_{\mathbf{k}_{\parallel},\mathbf{k}^v_{\parallel}} P^{vf\xi_v}_{\mathbf{k}^v_{\parallel},\mathbf{k}} \rangle |_{scatt} \nonumber \\
&+\frac{1}{i\hbar}\sum_{\mathbf{q}} V^{cv\xi_c\xi_v}_{\mathbf{q}} ~ \left(\rho^{v\xi_v}_{\mathbf{k}^v_{\parallel}+\mathbf{q}}-\rho^{c\xi_c}_{\mathbf{k}_{\parallel}+\mathbf{q}}\right) \delta\langle P^{\dagger\xi_c\xi_v}_{\mathbf{k}_{\parallel}+\mathbf{q},\mathbf{k}^v_{\parallel}+\mathbf{q}} P^{vf\xi_v}_{\mathbf{k}^v_{\parallel},\mathbf{k}} \rangle -i \Omega^{fc\xi_c}_{\mathbf{k}} ~ \delta\langle P^{\dagger\xi_c\xi_v}_{\mathbf{k}_{\parallel},\mathbf{k}^v_{\parallel}}P^{vc\xi_v\xi_c}_{\mathbf{k}^v_{\parallel},\mathbf{k}_{\parallel}}\rangle .\label{eq:eh_photo}
\end{align}
\end{widetext}
The source term of Eq. \eqref{eq:eh_photo} are TMDC-interband Coulomb correlations $\delta\langle P^{\dagger\xi_c\xi_v}_{\mathbf{k}_{\parallel},\mathbf{k}^v_{\parallel}}P^{vc\xi_v\xi_c}_{\mathbf{k}^v_{\parallel},\mathbf{k}_{\parallel}}\rangle$. The vacuum electron induced Pauli blocking of the transition Eq. \eqref{eq:eh_photo} is assumed to be small compared to the electron-hole population and is therefore neglected. For the TMDC interband transitions also occuring in Eq. \eqref{eq:rhof_cc} we obtain:
\begin{align}
\frac{d}{dt}P^{\xi_v\xi_c}_{\mathbf{k}_{\parallel}}&=\frac{1}{i\hbar}\left(\varepsilon^{c\xi_c}_{\mathbf{k}_{\parallel}}-\varepsilon^{v\xi_v}_{\mathbf{k}_{\parallel}}-\varepsilon_{vis}\right)P^{\xi_v\xi_c}_{\mathbf{k}_{\parallel}} \nonumber \\
&+\partial_t P^{\xi_v\xi_c}_{\mathbf{k}_{\parallel}}|_{scatt} -i\Omega^{cv\xi_c\xi_v}_{\mathbf{k}_{\parallel}}(\rho^{v\xi_v}_{\mathbf{k}_{\parallel}}-\rho^{c\xi_c}_{\mathbf{k}_{\parallel}}) \delta_{\xi_v,\xi_c} \nonumber \\
&-\frac{1}{i\hbar}\sum_{\mathbf{q}}V^{vc\xi_v\xi_c}_{\mathbf{q}} ~ (\rho^{v\xi_v}_{\mathbf{k}_{\parallel}}-\rho^{c\xi_c}_{\mathbf{k}_{\parallel}}) P^{\xi_v\xi_c}_{\mathbf{k}_{\parallel}+\mathbf{q}} .\label{eq:p_eh}
\end{align}
The attractive Coulomb interaction $V_{\mathbf{q}}^{vc\xi_v\xi_c}$ leads to a renormalization of the Rabi frequency of the exciting field \cite{Kira2006} and is treated by the Wannier equation after a coordinate transformation into the exciton basis. Since only the edges of the hexagonal Brillouin zone are optically excited and only momentum vertical transitions valid, the electron and hole valley of the TMDC interband polarization is restricted to $K$ or $K'$, respectively.

To transfer the equations to the excitonic basis \cite{Kochbuch,Katsch2018}, we introduce the center of mass momentum $\mathbf{Q}=\mathbf{k}^c_{\parallel}-\mathbf{k}^v_{\parallel}$ and the relative momentum $\mathbf{q}_{\parallel}=\alpha^{\xi_c}_{\xi_v}\mathbf{k}^v_{\parallel}+\beta^{\xi_c}_{\xi_v}\mathbf{k}^c_{\parallel}$ with the mass factors $\alpha^{\xi_c}_{\xi_v}=m_h^{\xi_v}/(m_e^{\xi_c}+m_h^{\xi_v})$ and $\beta^{\xi_c}_{\xi_v}=m_e^{\xi_c}/(m_h^{\xi_v}+m_e^{\xi_c})$. For equal electron and hole momenta the relative momentum equals the electron momentum. In this basis, we exploit the Wannier equation \cite{Kira2006,Axt2004,Kochbuch}:
\begin{align}
\frac{\hbar^2\mathbf{q}_{\parallel}^2}{2m}\varphi^{\xi_v\xi_c}_{\mu,\mathbf{q}_{\parallel}}-\sum_{\mathbf{k}_{\parallel}}V^{vc\xi_v\xi_c}_{\mathbf{k}_{\parallel}}\varphi^{\xi_v\xi_c}_{\mu,\mathbf{q_{\parallel}+k_{\parallel}}}=E_B^{\mu\xi_v\xi_c}\varphi^{\xi_v\xi_c}_{\mu,\mathbf{q}_{\parallel}} ,
\end{align}
to access the exciton wave function $\varphi^{\xi_v\xi_c}_{\mu,\mathbf{q}_{\parallel}}$ and binding energy $E_B^{\mu\xi_v\xi_c}$ of the exciton state $\mu$. The reduced mass of the exciton is denoted by $m$. By expanding the excitonic polarization into solutions of the Wannier equation $P^{\xi_v\xi_c}_{\mathbf{q_{\parallel},Q}}=\sum_{\mu}\varphi^{\xi_v\xi_c}_{\mu,\mathbf{q}_{\parallel}}P^{\xi_v\xi_c}_{\mu,\mathbf{Q}}$ the relative coordinate $\mathbf{q}_{\parallel}$ can be eliminated and the exciton dynamics is described by the center of mass momentum $\mathbf{Q}$. Using this expansion we provide all equations \eqref{eq:rhof_cc}-\eqref{eq:p_eh} in the exciton basis.

In the limit of negligible Pauli blocking, we obtain for the excitonic transition Eq. \eqref{eq:p_eh}:
\begin{align}
\frac{d}{dt} P^{\xi_v\xi_c}_{\mu,\mathbf{Q}}&=\frac{1}{i\hbar}\left(E^{\xi_v\xi_c}_{\mu,\mathbf{Q}}-\varepsilon_{vis}\right) P^{\xi_v\xi_c}_{\mu,\mathbf{Q}} + \partial_t P^{\xi_v\xi_c}_{\mu,\mathbf{Q}}\vert_{scatt} \nonumber \\
&-i\Omega^{\xi_v\xi_c}_{\mu}(t)~\delta^{\xi_v,\xi_c}_{\mathbf{Q},0} \label{eq:p_exc}
\end{align}
with the excitonic Rabi frequency $\Omega^{\xi_v\xi_c}_{\mu}(t)$ and the exciton kinetic energy, which reads $E^{\xi_v\xi_c}_{\mu,\mathbf{Q}}=\hbar^2\mathbf{Q}^2/2M^{\xi_v\xi_c}+E^{\xi_v\xi_c}_{\mu}$ where $M^{\xi_v\xi_c}=m_h^{\xi_v}+m_e^{\xi_c}$ denotes the exciton mass and $E^{\xi_v\xi_c}_{\mu}=E^{\xi_v\xi_c}_{gap}+E_B^{\mu\xi_v\xi_c}$ the exciton energy. Because of the assumed perpendicular excitation geometry only excitons with vanishing center of mass momentum couple to the light field. The phonon contribution leads to a dephasing of the excitonic transition and to the formation of incoherent excitons $N_{\mu,\mathbf{Q}}^{\xi_v\xi_c}=\delta\langle P^{\dagger\xi_c\xi_v}_{\mu,\mathbf{Q}}P^{\xi_v\xi_c}_{\mu,\mathbf{Q}}\rangle$. A detailed derivation of the exciton-phonon interaction and the microscopic computation of the dephasing can be found in Ref. \cite{Selig2016,Christiansen2017,Selig2018}. The used electron-phonon parameters stem from first-principle calculations described in Ref. \cite{Kaasbjerg2012,Kaasbjerg2013,Jin2014}. The phonon reservoir is treated in bath approximation following a Bose distribution.

For the photoemission signal Eq. \eqref{eq:rhof_cc}, now in the exciton basis, we obtain:
\begin{widetext}
\begin{align}
\frac{d}{dt} \rho^f_{\mathbf{k}}&=-2\im\left(\Omega^{vf\xi_v}_{\mathbf{k}} P^{vf\xi_v}_{\mathbf{k_{\parallel},k}} + \sum_{\mu} \Omega^{cf\xi_c-}_{\mu,\mathbf{k}} P^{*\xi_c\xi_v}_{\mu,0}P^{vf\xi_v}_{\mathbf{k}_{\parallel},\mathbf{k}} + \sum_{\mu,\xi_v,\mathbf{Q}} \Omega^{cf\xi_c}_{\mu,\mathbf{k,Q}} \delta\langle P^{\dagger\xi_c\xi_v}_{\mu,\mathbf{Q}}P^{vf\xi_v}_{\mathbf{k}_{\parallel}-\mathbf{Q},\mathbf{k}}\rangle \right) \label{eq:rhof_exc}
\end{align}
\end{widetext}
with the coupling elements $\Omega^{cf\xi_c-}_{\mu,\mathbf{k}}=\Omega^{cf\xi_c}_{\mathbf{k}} \varphi^{*\xi_c\xi_v}_{\mu,\mathbf{k}_{\parallel}}e^{-\frac{1}{i\hbar}\varepsilon_{vis}t}$ and $\Omega^{cf\xi_c}_{\mu,\mathbf{k,Q}}=\Omega^{cf\xi_c}_{\mathbf{k}} \varphi^{*\xi_c\xi_v}_{\mu,\mathbf{k}_{\parallel}-\alpha^{\xi_c}_{\xi_v}\mathbf{Q}}$. Now, we see that the photoemission of conduction band electrons occur always via excitonic states. The TMDC interband Coulomb interaction couples through the Fock term the coherently driven excitonic transition $P^{*\xi_c\xi_v}_{\mu,0}$ with the transition between valence band and vacuum (coherent source only present for coherent excitons and XUV-field) and, most importantly, as a source of Coulomb correlations, beyond the Hartree-Fock approximation we identify the exciton-assisted photoemission transition $\delta\langle P^{\dagger\xi_c\xi_v}_{\mu,\mathbf{Q}}P^{vf\xi_v}_{\mathbf{k}_{\parallel}-\mathbf{Q},\mathbf{k}}\rangle$:
\begin{table}[b]
\caption{\label{tab:electron_param}
Band structure parameters for WSe$_2$ obtained from first principle calculations \cite{Kormanyos2015,Guo2016} and following excitonic parameters (restricted to spin up).
}
\begin{ruledtabular}
\begin{tabular}{ l c c r }
\textrm{Param.}&
\textrm{}&
\textrm{Param.}&
\textrm{} \\
\colrule
$m_e^{K}/m_0$ & 0.29 & $M_{KK}/m_0$ & 0.65 \\
$m_e^{\Lambda}/m_0$ & 0.56 & $M_{K\Lambda}/m_0$ & 0.92 \\
$m_h^{K}/m_0$ & 0.36 & $M_{KK'}/m_0$ & 0.76 \\
$E_{gap}^{KK}/\text{eV}$ & 2.08 & $E_{B1s}^{KK}/\text{eV}$ & 0.23 \\
$E_{gap}^{K\Lambda}/\text{eV}$ & 2.075 & $E_{B1s}^{K\Lambda}/\text{eV}$ & 0.26 \\
$E_{gap}^{KK'}/\text{eV}$ & 2.057 & $E_{B1s}^{KK'}/\text{eV}$ & 0.25 \\
$E_{gap}^{K\Lambda'}/\text{eV}$ & 2.208 & $E_{B1s}^{K\Lambda'}/\text{eV}$ & 0.28 \\
$E_{ion}/\text{eV}$ & 5.17 & $\epsilon_{SiO_2}$ & 3.9
\end{tabular}
\end{ruledtabular}
\end{table}
\begin{figure*}[t!]
  \begin{center}
     \includegraphics[width=\linewidth]{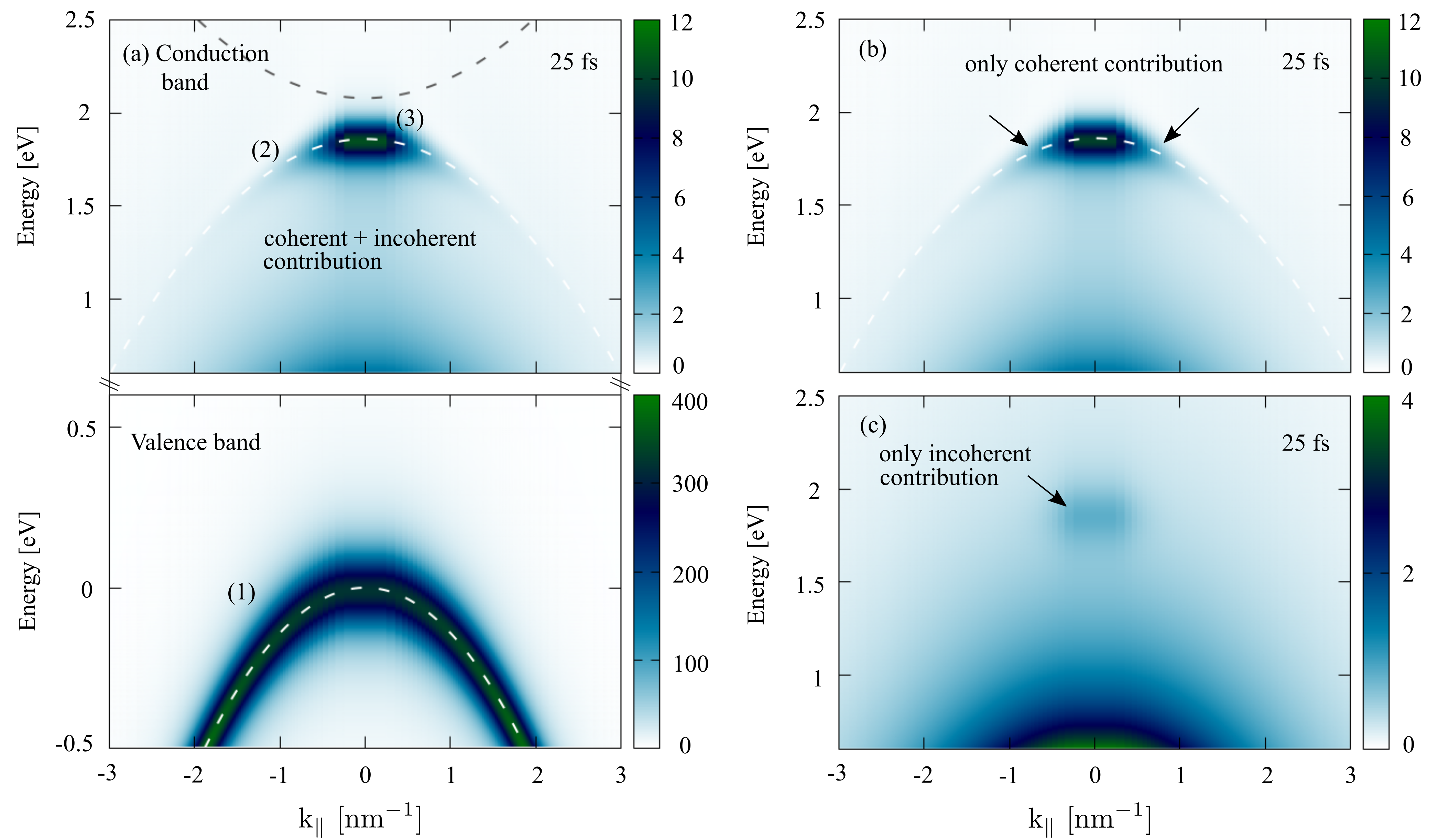}
   \end{center}
    \caption{The simulated photoemission signal of monolayer WSe$_2$ at \unit[77]{K} for the time delay of \unit[25]{fs}. (1), (2) and (3) refer to different aspects of the trARPES signal, which are discussed in Sec. \eqref{sec:ARPES}. (a) Conduction band region showing an excitonic signal stemming from coherent and incoherent $KK$-excitons at the exciton energy. The shadow of the valence band at the exciton energy reflects the presence of the coherent exciton. The grey dashed lines indicate the conduction band at the free-particle band gap and the white dashed line sketches the valence band dispersion at the exciton energy. At \unit[0]{eV} the fully occupied valence band is visible. (b) Coherent contribution to the trARPES signal in the conduction band region. (c) Incoherent contribution to the trARPES signal in the conduction band region.}
  \label{fig:coherence_arpes}
\end{figure*}
\begin{align}
\frac{d}{dt} \delta\langle P^{\dagger\xi_c\xi_v}_{\mu,\mathbf{Q}}P^{vf\xi_v}_{\mathbf{k}_{\parallel}-\mathbf{Q},\mathbf{k}}\rangle &= \frac{1}{i\hbar}\Delta\varepsilon_{\mathbf{k_{\parallel},Q},k_z} \delta\langle P^{\dagger\xi_c\xi_v}_{\mu,\mathbf{Q}}P^{vf\xi_v}_{\mathbf{k}_{\parallel}-\mathbf{Q},\mathbf{k}}\rangle \nonumber \\
&+\partial_t \delta\langle P^{\dagger\xi_c\xi_v}_{\mu,\mathbf{Q}}P^{vf\xi_v}_{\mathbf{k}_{\parallel}-\mathbf{Q},\mathbf{k}}\rangle\vert_{scatt} \nonumber \\
&- i\sum_{\lambda} \tilde{\Omega}^{fc\xi_v\xi_c}_{\mu,\lambda,\mathbf{k}_{\parallel},\mathbf{Q}} N^{\xi_v\xi_c}_{\lambda,\mathbf{Q}} \label{eq:Pcf_renorm}
\end{align}
with $\Delta\varepsilon_{\mathbf{k_{\parallel},Q},k_z}=\varepsilon^f_{\mathbf{k}}-\varepsilon^{v\xi_v}_{\mathbf{k}_{\parallel}-\mathbf{Q}}-E_{\mu,\mathbf{Q}}^{\xi_v\xi_c}-\varepsilon_{xuv}$ and $\tilde{\Omega}^{fc\xi_v\xi_c}_{\mu,\lambda,\mathbf{k},\mathbf{Q}}=\Omega^{fc\xi_c}_{\mathbf{k}} |\varphi^{\xi_v\xi_c}_{\lambda,\mathbf{k}_{\parallel}-\alpha^{\xi_c}_{\xi_v}\mathbf{Q}}|^2 \varphi^{\xi_v\xi_c}_{\mu,\mathbf{k}_{\parallel}-\alpha^{\xi_c}_{\xi_v}\mathbf{Q}}$. The oscillation energy of the TMDC exciton-assisted valence band-vacuum transition carries the exciton energy, is driven by the incoherent exciton density and carries therefore the information about the bound TMDC-excitons and their incoherent scattering dynamics. For completeness we provide also the equation of motion for $P^{vf\xi_v}_{\mathbf{k_{\parallel},k}}$ occuring in Eq. \eqref{eq:rhof_exc}:
\begin{align}
\frac{d}{dt} P^{vf\xi_v}_{\mathbf{k_{\parallel},k}}&=\frac{1}{i\hbar}\left(\varepsilon^f_{\mathbf{k}}-\varepsilon^{v\xi_v}_{\mathbf{k}_{\parallel}}-\varepsilon_{xuv}\right) P^{vf\xi_v}_{\mathbf{k_{\parallel},k}} +\partial_t P^{vf\xi_v}_{\mathbf{k_{\parallel},k}}\vert_{scatt} \nonumber \\
&-i\Omega^{fv\xi_v}_{\mathbf{k}}\rho^{v\xi_v}_{\mathbf{k}_{\parallel}} -i\sum_{\mu}\Omega^{fc\xi_c+}_{\mu,\mathbf{k}} P^{\xi_v\xi_c}_{\mu,0} \label{eq:Pvf_exc}
\end{align}
with $\Omega^{fc\xi_c+}_{\mu,\mathbf{k}}=\Omega^{fc\xi_c}_{\mathbf{k}}\varphi^{\xi_v\xi_c}_{\mu,\mathbf{k}_{\parallel}}~e^{\frac{1}{i\hbar}\varepsilon_{vis}t}$.

Next, we investigate the photoemission signal Eq. \eqref{eq:observable} by analyzing its individual contributions. As a first attack, we focus on the lowest lying 1s A state $P_{\mu,0}^{\xi_v\xi_c}\rightarrow P^{\xi_v\xi_c}_{0}$ and $N^{\xi_v\xi_c}_{\lambda,\mathbf{Q}}\rightarrow N^{\xi_v\xi_c}_{\mathbf{Q}}$, justified by the large 1$s$-2$s$ separation in comparison to the thermal energy introduced by exciton-phonon scattering mediated thermalization \cite{Brem2018,Brem2019}

\section{Time-resolved ARPES} \label{sec:ARPES}

For the numerical evaluation of Eq. \eqref{eq:observable} we choose a VIS pulse, which excites resonantly the 1s A exciton (\unit[1.86]{eV}). The subsequent XUV probe pulse has an excitation energy of \unit[20]{eV} for the photoemission. The pump and probe pulse field intensity width is \unit[35]{fs} and \unit[20]{fs}, respectively. The calculations are performed for an exemplary temperature of \unit[77]{K} in the exemplary material WSe$_2$ on a quartz substrate. For higher temperature we can expect a similar behaviour in WSe$_2$ except faster time scales due to a higher phonon occupation and more efficient exciton-phonon scattering \cite{Selig2018}. Table \ref{tab:electron_param} summarizes the used electronic and excitonic parameters.

\subsection{Excitonic signal at pulse overlap}

To address the coherent signals at pulse overlap first, Fig. \ref{fig:coherence_arpes} displays the result for monolayer WSe$_2$ with a pulse delay of \unit[25]{fs} for the conduction and valence band region. Note that the $k_{\parallel}$-axis has been shifted on top of the $K$-point. We discuss all observed features (1-3) in Fig. \ref{fig:coherence_arpes} as follows:

(1) First, in Fig. \ref{fig:coherence_arpes} (a) at \unit[0]{eV}, the valence band electron dispersion can be recognized. This contribution results from the valence band-vacuum transition $P^{vf}_{\mathbf{k_{\parallel},k}}$ in Eq. \eqref{eq:rhof_exc}. In principle, at the band maximum a reduced trARPES signal reflects the optically excited hole distribution building up 1s excitons. However, in the low-excitation regime considered here, this contribution is vanishing small and cannot be seen in the plot.

(2) Second, in Fig. \ref{fig:coherence_arpes} (a) and (b), we find a coherent excitonic signal at the exciton energy, which features a strong exciton contribution and a weak shadow of the valence band dispersion highlighted by the white dashed line at the exciton energy. To clarify the coherent contribution to the trARPES signal Fig. \ref{fig:coherence_arpes} (b) displays the result for the conduction band region when setting the incoherent excitonic densities to zero $N_{\mathbf{Q}}^{\xi_v\xi_c}=0$. Consequently, only the first two terms of Eq. \eqref{eq:rhof_exc} are non-vanishing. This coherent feature results from the coupling of the optically induced coherent excitonic transition $P_{0}^{\xi_v\xi_v}$ ($\mathbf{Q}=0$ in the coherent limit) and the photoemission transition $P^{vf\xi_v}_{\mathbf{k_{\parallel},k}}$ between valence band and vacuum states, cf. Eq. \eqref{eq:rhof_exc} second term. Since the optically excited excitonic transition exists only for vanishing center of mass momentum, this shadow of the valence band is only visible at the $K$-point. The valence band shadow is illustrated by arrows in Fig. \ref{fig:coherence_arpes} (b). The decrease of the valence band shadow along the in-plane momentum is determined by the wave number decay of the exciton wave function $|\varphi_{\mathbf{q}_{\parallel}}|^2$, as it is obvious from the coupling element Eq. \eqref{eq:rhof_exc}. \textit{Therefore, from Fig. \ref{fig:coherence_arpes}, we conclude that in the ultrafast coherent limit trARPES is a technique to image the exciton wave function and measure the exciton Bohr radius  in momentum space \cite{Rustagi2018}.} To obtain more analytical insights into the dominating excitonic signal, we can formally integrate the equations \eqref{eq:p_exc}, \eqref{eq:rhof_exc} and \eqref{eq:Pvf_exc} assuming  exponentially shaped pulses of the form $\exp(-|t-\tau|/\sigma)$ with width $\sigma$. From the TMDC interband Coulomb contribution we obtain a resonance of the signal, Eq. \ref{eq:observable}, at $I_{\mathbf{k}_{\parallel},\varepsilon^f_{\mathbf{k}}}\propto |\varphi^{\mathstrut}_{\mathbf{k}_{\parallel}}|^2\rho^v_{\mathbf{k}_{\parallel}}\delta\left(\varepsilon^f_{\mathbf{k}}-\varepsilon_{xuv}-E^{1s}_{0}-\varepsilon^v_{\mathbf{k}_{\parallel}}\right)$, where we assumed vanishing dephasing rates to use strict energy conservation. Clearly, the trARPES signal scales in $k_{\parallel}$ with the exciton wave function $|\varphi^{\mathstrut}_{\mathbf{k}_{\parallel}}|^2$.

\begin{figure}[t!]
  \begin{center}
     \includegraphics[width=\linewidth]{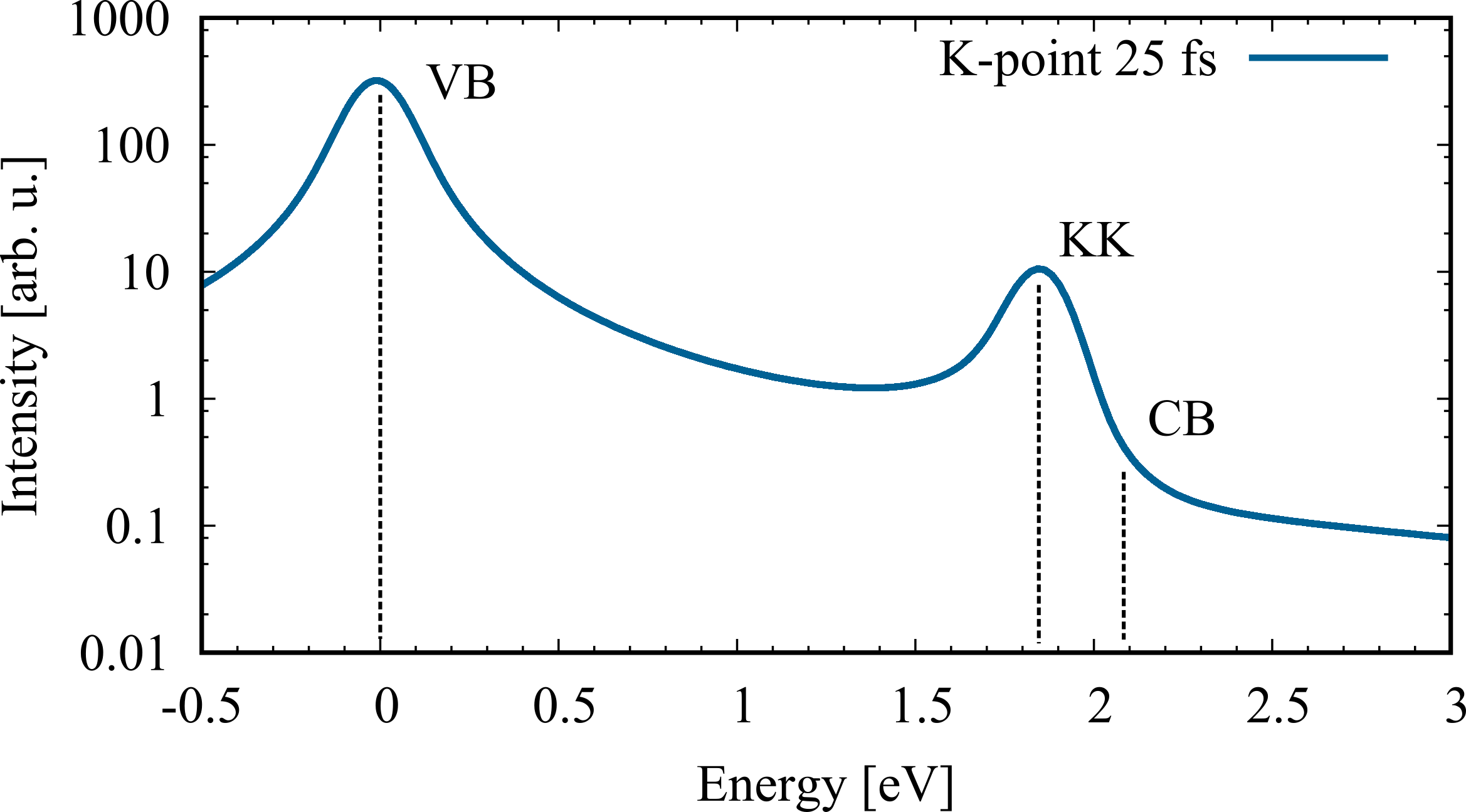}
   \end{center}
    \caption{Energy distribution curve in the low excitation regime for WSe$_2$ at the $K$-point \unit[25]{fs} after optical excitation. Besides the valence band we find a peak at the exciton energy from the photoemission of conduction band electrons, forming here mainly, coherent but also incoherent $KK$-excitons. Note the logarithmic scale.}
  \label{fig:EDC_50}
\end{figure}

(3) Third, in Fig. \ref{fig:coherence_arpes} (a) and (c), incoherent $KK$-excitons, localized at the exciton energy, also generate an excitonic contribution in trARPES. The corresponding signal in Fig. \ref{fig:coherence_arpes} (a) is, as the coherent signal (2), also down-shifted by the binding energy with respect to the single-particle band gap (dashed line above \unit[2]{eV}). Figure \ref{fig:coherence_arpes} (c) shows the trARPES result for the incoherent limit with vanishing excitonic transition $P^{\xi_v\xi_c}_0=0$. The exciton population $N_{\mathbf{Q}}^{\xi_v\xi_c}$, which is the source of this signal, is determined by the exciton-phonon induced scattering transfer of optically excited coherent to incoherent excitons \cite{Thranhardt2000}. The corresponding equation for $N_{\mathbf{Q}}^{\xi_v\xi_c}$ can be found in Ref. \cite{Selig2018}. We consider the high symmetry points $K,K',\Lambda$ and $\Lambda'$ to compute the exciton formation and relaxation throughout the excitonic Brillouin zone. Therefore, the correlated two-particle quantity $\delta\langle P^{\dagger\xi_c\xi_v}_{\mathbf{Q}}P^{vf\xi_v}_{\mathbf{k}_{\parallel}-\mathbf{Q},\mathbf{k}}\rangle$ in Eq. \eqref{eq:rhof_exc} contains the information about the exciton dynamics, namely formation and relaxation. Obviously, the correlated exciton-assisted photoemission quantity determines the signal by a convolution of the exciton wave function and distribution, cf. Eq. \eqref{eq:rhof_exc} and \eqref{eq:Pcf_renorm}, when inserting the definitions for $\Omega^{fc\xi_c}_{\mu,\mathbf{k,Q}}$ and $\tilde{\Omega}^{fc\xi_v\xi_c}_{\mu,\lambda,\mathbf{k},\mathbf{Q}}$ right below. Together with the sum over the center of mass momentum along the exciton dispersion the trARPES signal lubricates and does not display the exact valence band shadow at the exciton energy.

We conclude that the excitonic signal (2) in Fig. \ref{fig:coherence_arpes} has two excitonic contributions, a coherent and an incoherent. But the main contribution for such short delay times stems from the coherent exciton since the incoherent excitons have first to be build up through phonon-induced dephasing from the excitonic transition \cite{Selig2018}.

Figure \ref{fig:EDC_50} shows the trARPES intensity for a fixed in-plane momentum $k_{\parallel}$, typically referred to as energy distribution curve (EDC), at the $K$-point. As expected from Fig. \ref{fig:coherence_arpes}, we observe two peaks, the first displaying the valence band and the second showing the exciton. The energetic position of the conduction band is also depicted as dashed line for comparison. The shown result is valid for the low excitation regime, reflected by the weak exciton signal compared to the valence band. Expanding the photoemission term of valence band electron in Eq. \eqref{eq:rhof_exc} by inserting a unit operator $\sum_{\xi_c,\mathbf{k}^c_{\parallel}}c^{\dagger\xi_c}_{\mathbf{k}^c_{\parallel}}c^{\xi_c}_{\mathbf{k}^c_{\parallel}}$ leads to
\begin{figure*}[t!]
  \begin{center}
	  \includegraphics[width=\linewidth]{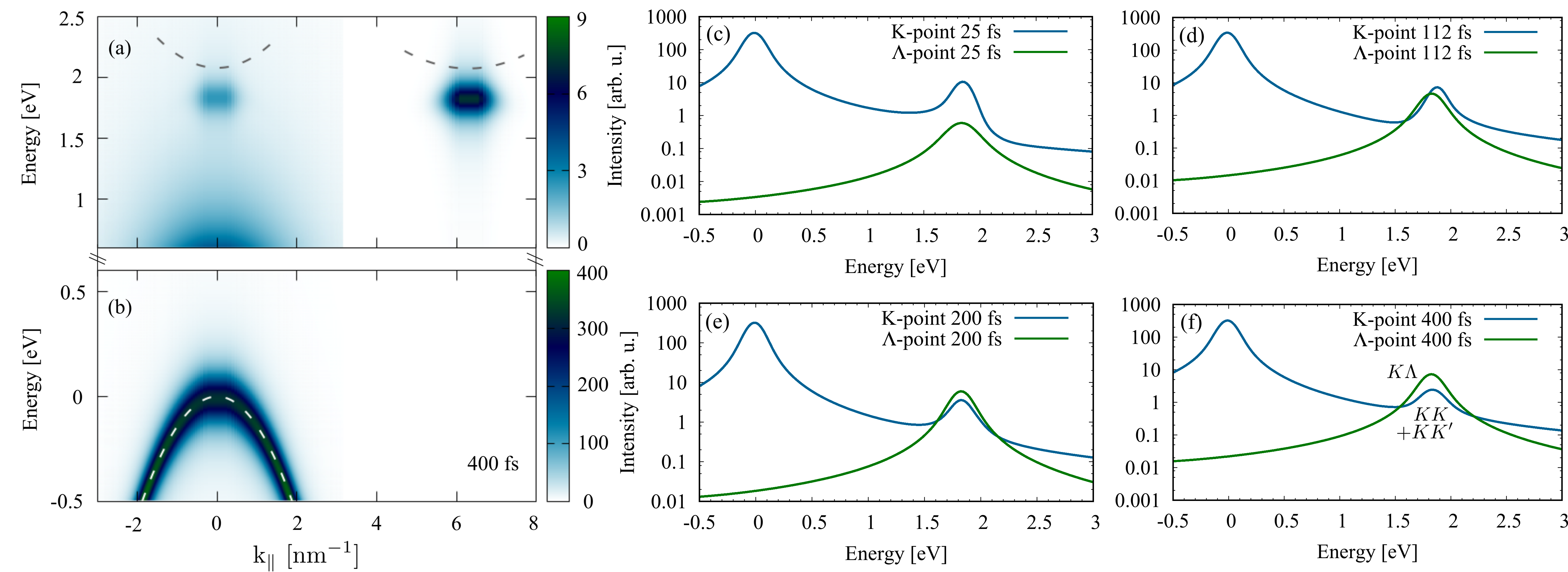}
   \end{center}
    \caption{The simulated photoemission signal of monolayer WSe$_2$ at \unit[77]{K} for the time delay of \unit[400]{fs} and EDCs for different time delays. (a) We observe the relaxation of electrons into the $\Lambda$-valley, forming momentum-indirect and energetically more favourable $K\Lambda$-excitons visible as excitonic peak below the $\Lambda$-conduction band (dashed) at $k_{\parallel}\approx\unit[6]{nm^{-1}}$. An excitonic signal is remaining at the $K$-point due to the formation of also energetically more favourable momentum indirect $KK'$-excitons. (b) Still shows the almost unpertubed valence band. (c)-(f) Energy distribution curves for time delays of \unit[25]{fs}, \unit[112]{fs}, \unit[200]{fs} and \unit[400]{fs} showing the relaxation and thermalization of excitons into momentum-indirect excitonic states.}
  \label{fig:incoherent_arpes}
\end{figure*}
\begin{align}
P^{vf\xi_v}_{\mathbf{k_{\parallel},k}}&=P^{vf\xi_v}_{\mathbf{k_{\parallel},k}}-\sum_{\mu,\xi_c}\varphi^{\xi_v\xi_c}_{\mathbf{k}_{\parallel}}P^{\xi_v\xi_c}_{\mu,0}P^{cf\xi_c}_{\mathbf{k_{\parallel},k}} \nonumber \\
&-\sum_{\mu,\xi_c,\mathbf{Q}}\varphi^{\xi_v\xi_c}_{\mu,\mathbf{k}_{\parallel}+\beta^{\xi_c}_{\xi_v}\mathbf{Q}}\delta\langle P^{\xi_v\xi_c}_{\mu,\mathbf{Q}}P^{cf\xi_c}_{\mathbf{k_{\parallel}+Q,k}}\rangle ,
\end{align}
resulting in excitonic corrections to the dominating electronic contribution of the valence band $P^{vf\xi_v}_{\mathbf{k_{\parallel},k}}$. The appearing correlated two-particle quantity, which is driven by the incoherent excitons, would lead to a weak excitonic satellite at the exciton binding energy above the valence band \cite{Steinhoff2017}. However, as already discussed, we focus on the low excitation regime with $N^{\xi_v\xi_c}_{\mathbf{Q}}\ll 1$ and neglect these corrections.

\subsection{Exciton signals at large pulse delays}

So far we investigated only short time delays and focused on the optically excited $K$-point. Figure \ref{fig:incoherent_arpes} (a) and (b) displays the result for a time delay much larger than the typical exciton-phonon scattering times, here taken as \unit[400]{fs}. Because of the exciton-phonon scattering, side valleys, as the $\Lambda$-valleys ($k_{\parallel}\approx\unit[6]{nm^{-1}}$) or $K'$-valley get populated. The reason is that the formation of momentum-indirect excitons, with a hole at the $K$-point and an electron at the $\Lambda$- or $K'$-point, are energetically more favourable than the direct $KK$-excitons. The momentum-indirect incoherent $K\Lambda$-excitons can directly be observed at the $K\Lambda$-exciton energy below the conduction band at the $\Lambda$-point ($k_{\parallel}\approx\unit[6]{nm^{-1}}$) in time-resolved ARPES. Note, that we show the trARPES signal as a function of the electronic wave number's absolute value. Therefore, all three $\Lambda$-valleys of the 1st Brillouin zone contribute to the same trARPES signal at $k_{\parallel}=\Lambda$. The $KK'$-excitons have as center of mass momentum $\mathbf{Q=K}$. Therefore, in trARPES, where we investigate the signal as function of the absolute value of the electronic momentum $k_{\parallel}$, the $KK'$-exciton overlaps in momentum space with the $KK$-exciton and its contribution to the trARPES signal lies at the $K$-point. In contrast to very short delay times, at large delays the signal is determined by the presence of incoherent excitons and incoherent exciton-phonon scattering can be investigated. Note that the exciton dynamics fulfil a detailed balance between coherent excitons $|P_0^{\xi_v\xi_c}|^2$, incoherent excitons $N_{\mathbf{Q}}^{\xi_v\xi_c}$ and the radiative decay.

The relaxation into $K\Lambda$-states can be seen more clearly by investigating the energy distribution curves in Fig. \ref{fig:incoherent_arpes} for different consecutive pump-probe delay times around \unit[1.8]{eV}. Within the approximately first \unit[100]{fs} we observe the formation of incoherent excitons, cf. Fig. \ref{fig:incoherent_arpes} (c), (d) and then the subsequent thermalization, cf. Fig. \ref{fig:incoherent_arpes} (e), (f).

A different picture can be found in molybdenum-based TMDCs. Here, due to our calculations the $KK$-exciton state is the global minimum. Consequently, only intravalley exciton-phonon scattering, mainly with acoustic phonons, takes place leading to a thermalization of excitons \cite{Selig2018} and an trARPES signal at the $K$-point.

\hspace{-10cm}
\section{Conclusion}

In conclusion, we developed a theory of excitonic time-resolved ARPES in the low-excitation limit. We find that interband electron-hole Coulomb interaction strongly influences the photoemission spectrum and therefore we observe excitonic features in photoemission. We have demonstrated that the photoemission through the optically injected excitonic transitions leads to unintuitive signals, namely a shadw of the valence band at the excitonic energy. Additionally, we reveal that trARPES is able to probe the exciton dynamics and the exciton Bohr radius in TMDCs. We expect that our results are scalable to other 2D structures such as van der Waals heterostructures, where more complex exciton dynamics takes place namely through energy or charge transfer of the excitons from one layer to another \cite{Selig2019,Ovesen2019,Aeschlimann2019}.

\vspace{5mm}
We acknowledge financial support from the Deutsche Forschungsgemeinschaft (DFG) through SFB 951 (A.K., R.E.,  M.S., D.C.) Projektnummer 182087777. This project has also received funding from the European Unions Horizon 2020 research and innovation programme under Grant Agreement No. 785219 (Graphene Flagship, E.M.) and No. 734690 (SONAR, A.K.). E.M. also acknowledges support from the Swedish Research Council (VR). R.E. acknowledges funding from the European Research Council (ERC) under the European Union’s Horizon 2020 research and innovation program (Grant Agreement No. ERC-2015- CoG-682843). D.C. thanks the school of nanophotonics (SFB 787) for support. Finally, we thank Florian Katsch, Sandra Kuhn and Marten Richter (TU Berlin) for valuable discussions.

\appendix

\section{Wave functions} \label{app:fct}
The goal is to find a description of the electronic states in the confinement potential of a two-dimensional semiconductor including its vacuum states. The Schr\"odinger equation for an ideal 2d confinement potential at $z=0$ with height $V_0$ and the lattice periodic potential $V(\mathbf{r}_{\parallel})$ reads:
\begin{align}
E\xi(\mathbf{r})&=-\frac{\hbar^2}{2m}\left(\nabla^2_{\parallel}+\frac{\partial^2}{\partial z^2}\right)\xi(\mathbf{r}) \nonumber \\
&+\left(V(\mathbf{r_{\parallel}})\delta(z)-V_0\delta(z)\right)\xi(\mathbf{r})
\end{align}
where we assumed an ideal delta-like confinement potential.

To simplify the problem we drop the delta function, which limits the lattice potential to $z=0$ that the potential decompose in a parallel and transversal part. To obtain the single-particle wave functions we perform a nanostructure-envelope formalism inspired ansatz, where the wave function is expressed as product of an envelope wave function $\xi^{\lambda}(\mathbf{r})$ and the lattice periodic Bloch function $u_{\mathbf{k}}^{\lambda}(\mathbf{r})$
\begin{align}
\psi^{\lambda}_{\mathbf{k}}(\mathbf{r})=\xi^{\lambda}(\mathbf{r})u_{\mathbf{k}}^{\lambda}(\mathbf{r})
\end{align}
with $\lambda\in\{f,c,v\}$, where the envelope $\xi(\mathbf{r})$ is obtained from the Schr\"odinger equation. The solution reads
\begin{align}
\xi^{\lambda}=
   \begin{cases}
     \frac{1}{\sqrt{S}}~ e^{i\mathbf{k^{\lambda}_{\parallel}r_{\parallel}}}\left(e^{-ik_z^{\lambda}z}+\frac{1}{i\beta k_z^{\lambda}-1}e^{ik_z^{\lambda}z}\right) & \lambda=f \\
     \frac{\sqrt{k_z^{\lambda}}}{\sqrt{S}}~ e^{i\mathbf{k^{\lambda}r_{\parallel}}}~e^{-k_z^{\lambda}z} & \lambda=c \lor v     
   \end{cases}
\end{align}
with $\beta=\hbar^2/(mV_0)$. The bound solutions ($\lambda=c/v$) consist of a plane wave in parallel direction to the confinement function. Due to the exponential decay of the wave function into the vacuum and the semiconductor layer, the bound solution resemble closely to surface states. The unbound solution ($\lambda=f$) consits in parallel direction also of plane waves. Perpendicular to the surface we use the scattering solution of the plane wave at the confinement potential. Together with the lattice periodic function we obtain as electronic wave functions
\begin{align}
\psi^{c/v}_{\mathbf{k}_{\parallel}}(\mathbf{r})&=\sqrt{\frac{k_z^{c/v}}{S}}e^{i\mathbf{k}^{c/v}_{\parallel}\cdot\mathbf{r_{\parallel}}}~e^{-k_z^{c/v}z}~ u^{c/v}_{\mathbf{k}_{\parallel}}(\mathbf{r}) \label{WF0}\\
\psi^{f}_{\mathbf{k}}(\mathbf{r})&=\frac{1}{\sqrt{S}}e^{-i\mathbf{k}^f_{\parallel}\mathbf{r_{\parallel}}}\left(e^{-ik_z^{f}z}+\frac{1}{i\beta k_z^f-1}e^{ik_z^{f}z}\right)u^f_{\mathbf{k}}(\mathbf{r}). \label{WF}
\end{align}

\section{Dipole matrix elements} \label{app:optical}
With the electronic wave functions \eqref{WF0} and \eqref{WF} we can compute the optical matrix elements. For the computation we separate the slowly-varying envelope and cell-periodic parts. In $\mathbf{r}\cdot\mathbf{E}$-coupling the matrix elements read:
\begin{align}
d^{\lambda\lambda'}_{\mathbf{k,k'}}&=-e\int d^3r ~ \psi^{*}_{\lambda,\mathbf{k}}(\mathbf{r})\mathbf{E}(t)\cdot\mathbf{r}\psi_{\lambda',\mathbf{k}'}(\mathbf{r})
\end{align}
Since we assume that the electric field is constant over the unit cell, it is space independent and we can take it out of the integral. Inserting the wave functions and after shifting the integral into the first unit cell we obtain:
\begin{align}
d^{cv}_{\mathbf{k_{\parallel}}}&=-\frac{e}{S}\frac{\sqrt{k_z^c k_z^v}}{k_z^c + k_z^v}\cdot N\delta_{\mathbf{k},\mathbf{k}'}\cdot \mu^{cv}_{\sigma} \label{optcv} \\
d^{c/vf}_{\mathbf{k_{\parallel},k}}&=-ie\frac{N}{S}\sqrt{k_z^{c/v}}\left(\frac{1}{(\beta k_z^f+i)\cdot (k_z^f+i k_z^{c/v})} \right. \nonumber \\
&\hspace{2.5cm}\left. +\frac{1}{k_z^{c/v}+ik_z^f}\right) \times \nonumber \\
&\times\left(\int_{V}d^3r~u^{*c/v}_{\mathbf{k}_{\parallel}}(\mathbf{r})\nabla_{\mathbf{k}_{\parallel}} u^f_{\mathbf{k}}(\mathbf{r})+\delta_{c/v,f}\nabla_{\mathbf{k}_{\parallel}}\right).
\end{align}
$V$ denotes the unit cell volume, $S/N$ the unit cell surface and $\mu^{cv}_{\sigma}=\int_{V} d^3r ~ u_{\mathbf{k}_{\parallel}}^{*c}(\mathbf{r}) ~ \mathbf{r}\cdot\mathbf{e}_{\sigma} ~ u^v_{\mathbf{k}_{\parallel}}(\mathbf{r})$ the microscopic matrix element with the polarization vector $\mathbf{e}_{\sigma}$. For the transition element from the conduction band to the vacuum states, we used the identity $\mathbf{r}\exp(-i\mathbf{k\cdot r})=i\nabla_{\mathbf{k}}\exp(-i\mathbf{k\cdot r})$ leading to an intra- and interband contribution to the dipole element.

\hspace{-5cm}
\section{Coulomb matrix elements} \label{app:cc}
The Coulomb matrix element reads
\begin{align}
V^{\lambda_1,\lambda_2 ~ \mathbf{k_1,k_2}}_{\lambda_3,\lambda_4 ~ \mathbf{k_3,k_4}}=\sum_{\mathbf{q}} V_{\mathbf{q}} ~ \Gamma^{\lambda_1\mathbf{k}_1}_{\lambda_3\mathbf{k}_3}(\mathbf{q}) \Gamma^{\lambda_2\mathbf{k}_2}_{\lambda_4\mathbf{k}_4}(-\mathbf{q}) \label{cccv}
\end{align}
with
\begin{align}
\Gamma^{\lambda_1\mathbf{k}_1}_{\lambda_3\mathbf{k}_3}(\mathbf{q})=\int d^3r ~ \psi^*_{\lambda_1,\mathbf{k}_1}(\mathbf{r}) e^{i\mathbf{q\cdot r}} \psi_{\lambda_3,\mathbf{k}_3}(\mathbf{r}) \\
\Gamma^{\lambda_2\mathbf{k}_2}_{\lambda_4\mathbf{k}_4}(-\mathbf{q})=\int d^3r ~ \psi^*_{\lambda_2,\mathbf{k}_2}(\mathbf{r}) e^{-i\mathbf{q\cdot r}} \psi_{\lambda_4,\mathbf{k}_4}(\mathbf{r})
\end{align}
and $V_{\mathbf{q}}$ being the Fourier transform of the Coulomb potential. The integrals can be calculated by shifting the integral into the first elementary cell $\mathbf{r\rightarrow r+R}_n$ and summing over all unit cells. We find
\begin{widetext}
\begin{align}
V^{\lambda_1,\lambda_2 ~ \mathbf{k}_1,\mathbf{k}_2}_{\lambda_3,\lambda_4 ~ \mathbf{k}_3,\mathbf{k}_4}&=\mathcal{N}\left(\frac{S}{N}\right)^2\sum_{\mathbf{q}}V_{\mathbf{q}}\left(\delta_{\mathbf{k}_3-\mathbf{k}_1+\mathbf{q},0} ~ \delta_{\mathbf{k}_4-\mathbf{k}_2-\mathbf{q},0}\int dz ~  \zeta^*_{\lambda_1,k_{z}^1}\zeta^{\mathstrut}_{\lambda_3,k_{z}^3} \int dz ~ \zeta^*_{\lambda_2,k_{z}^2}\zeta^{\mathstrut}_{\lambda_4,k_{z}^4} \times \right. \nonumber \\
&\left.\times \int d^3r ~ u^*_{\lambda_1,\mathbf{k}_3+\mathbf{q}}u_{\lambda_3,\mathbf{k}_3}^{\mathstrut} \int d^3r ~ u^*_{\lambda_2,\mathbf{k}_2+\mathbf{q}}u_{\lambda_4,\mathbf{k}_4}^{\mathstrut}\right).
\end{align}
\end{widetext}
$\mathcal{N}$ denotes the normalization constant from the wave functions. The integral over the Bloch amplitudes can be treated in $\mathbf{k\cdot p}$-expansion with zeroth order describing intraband and first order interband interaction.

\bibliographystyle{apsrev4-1}

\end{document}